\documentclass[letterpaper]{article} 
\usepackage{aaai24}  
\usepackage{times}  
\usepackage{helvet}  
\usepackage{courier}  
\usepackage[hyphens]{url}  
\usepackage{graphicx} 
\def\UrlFont{\rm}  
\usepackage{natbib}  
\usepackage{caption} 
\usepackage{algorithm}
\usepackage{algorithmic}

\usepackage{newfloat}
\usepackage{listings}
\DeclareCaptionStyle{ruled}{labelfont=normalfont,labelsep=colon,strut=off} 
\floatstyle{ruled}
\newfloat{listing}{tb}{lst}{}
\floatname{listing}{Listing}
\title{Detecting and Triaging Spoofing using Temporal Convolutional Networks}
\author{
    Kaushalya Kularatnam,
    Tania Stathaki
}
\affiliations{
   
    Department of Electrical and Electronic Engineering, Imperial College London, UK\\
}

\usepackage{bibentry}

\begin{document}

\maketitle

\begin{abstract}
As algorithmic trading and electronic markets continue to transform the landscape of financial markets, detecting and deterring rogue
agents to maintain a fair and efficient marketplace is crucial. The explosion of large datasets and the continually changing tricks of the trade make it difficult to adapt to new market conditions and detect bad actors. To that end, we propose a framework that can be adapted easily to various problems in the space of detecting market manipulation. Our approach entails initially employing a labelling algorithm which we use to create a training set to learn a weakly supervised model to identify potentially suspicious sequences of order book states. The main goal here is to learn a representation of the order-book that can be used to easily compare future events. Subsequently, we posit the incorporation of expert assessment to scrutinize specific flagged order book states. In the event of an expert's unavailability, recourse is taken to the application of a more complex algorithm on the identified suspicious order book states. We then conduct a similarity search between any new representation of the order book against the expert labelled representations to rank the results of the weak learner. We show some preliminary results that are promising to explore further in this direction. 
\end{abstract}

\section{Introduction and Motivation}

The proliferation of algorithmic trading in modern financial markets brings about new challenges for the regulators and the regulated. Technological hurdles in processing vast amounts of data at very high speeds, identifying price manipulation occurring at micro-second level granularity and adapting to new market conditions and increasing number of players are some of the challenges faced by several entities. While technological shortcomings can be addressed, the process of identifying various forms of market manipulation is a highly challenging task. Accurately identifying market manipulation is crucial because false prices can have a detrimental impact on the price formation process. This distortion of prices, induced by bad agents, can create a chain of non-optimal trades and can affect other connected markets.

One such price manipulation technique is known as Spoofing. The financial crash of 2010 first brought this technique to prominence and has since been a concern for various regulators across the globe. Previous research on spoofing has generally followed two approaches: (i) Empirical analysis of spoofing cases and the accompanying deviation in market variables. e.g Handcrafted rules that attend to detect a large order by comparing it against historic averages \cite{Leeaun} (ii) Detection based on known cases of spoofing \cite{Qureshi}. There are also some early attempts of using labelled data, clustering and anomaly detection techniques to detect spoofing\cite{tradingtech}. In this work, we lay the foundations for learning a representation of spoofing using weak supervision, which has the potential to assist in detecting various forms of spoofing in stock markets. 

Spoofing is a form of market manipulation and involves placing non-bona fide orders on one side of the book misleading investors and algorithms into believing that there is excessive demand or supply for an asset. This in turn may incentivize investors into placing further orders which may artificially inflate or depress the price of the stock momentarily and possibly lock investors into an unfavourable price. The European regulator ESMA defines Layering and Spoofing as "submitting multiple orders often away from the inside on one side of the order book with the intention of executing a trade on the other side of the order book. Once that trade has taken place, the manipulative orders will be removed." \cite{mar}

\begin{figure}[h]
  \centering
  \includegraphics[width=\linewidth]{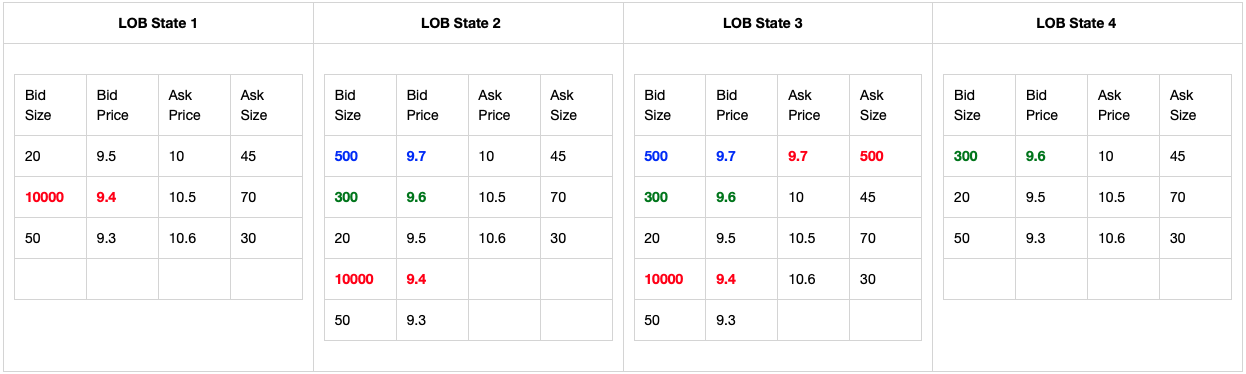}
  \caption{Eg. Spoofing the Limit Order Book. The four states of the order book do not always have to be consecutive and there can be multiple book updates in between these states}
\end{figure}

Spoofing can be very difficult to detect as it can be camouflaged among a high number of updates to a Limit Order Book (LOB). Not only do price manipulation activities such as spoofing affect the asset prices, if identified and prosecuted they involve hefty fines for the parties involved. In 2010, FINRA fined nine proprietary traders at Trillium Brokerage Services for spoofing. On June 25, the US Commission Futures Trading Commission (CFTC) fined Merrill Lynch Commodities \$25m for spoofing, manipulation, and attempted manipulation over a six year period \cite{cftcfine}. 

\section{Model and Training}

In order to be able to achieve our ultimate goal of using learned representations to identify similarities between various order book states, we need to determine if our target can be learned. With this in mind we conducted some initial experiments which show promising results:

\begin{itemize}
  \item We stack a series of market order book states together and processed them through a labelling algorithm that creates windowed time series data with labels.
  \item Subsequently, we conducted experiments with a validation set, whose labelling algorithm was slightly different to the training labelling algorithm.
  \item We show that we can predict the correct labels with 91\% accuracy.
  \item We propose that these highlighted potentially suspicious activities be labelled by an expert.  
  \item Additionally, we suggest learning a similarity measure between any new potentially suspicious states identified by our weak learner and the high confidence annotations we have from our experts.  
  \item We rank the results of the weak learner based on its proximity to the high confidence annotations. 
\end{itemize}

Our modelling task now involves a time-series, which is our MOB $x_0.....x_T$ and labels $y_0.....y_T$ where $y_i \in {0, 1, 2}$ where label $0$ has no classification, $1$ is buy side spoofing and $2$ is sell side spoofing. $x_0$ is a matrix $(h, w, d)$ where $h$ is 30, $w$ is 2, and $d$ is 2. Here $d_1$ will be $2d$ matrix with representing the quantity present on the book and $d_2$ will be a $2d$ matrix with the corresponding prices on the book. We stack our time series market order book data on top of each other for the desired number of frames (or timesteps) to predict an outcome. After this operation $x_0$ will be a matrix $n, h, w, d$ where $n$ is the number of frames and the other dimensions remain. Our model is causal, $y_t$ only depends on $x_{0:t}$ and not on $x_{t+1:T}$ - i.e information from the future is not injected into the past. In our modelling we use a temporal convolutional network (TCN) which has 128 filters with a kernel size $k = 2$ and dilations d of ${1, 2, 4, 8, 16, 32, 64}$. We use the swish activation function in our TCN layers as described in \cite{ramach2017searching} which worked better than ReLU for our predictions. 

\begin{figure}[h]
  \centering
  \includegraphics[width=\linewidth]{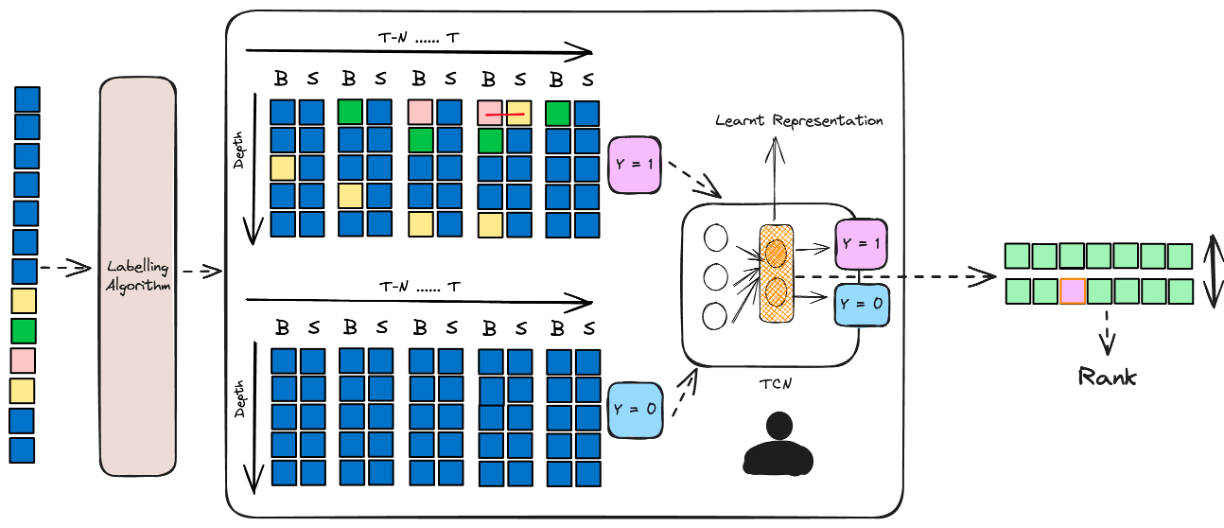}
  \caption{Framework we propose to detect Spoofing}
\end{figure}

\section{Preliminary Experiments}
Our dataset consisted of five stocks: \emph{SPY}, \emph{GOOG}, \emph{IBM}, \emph{VOD}, \emph{ABEO}, trading on five exchanges: \emph{NASDAQ}, \emph{EDGX}, \emph{BATS}, \emph{EDGA} and \emph{ARCA}. All data was capped to 30 levels from the top of the book. We also have a mix of liquid and illiquid instruments in our training, test and validation sets. 

The table can be read as 10 market order book states trained with cross entropy loss.  E=256 indicates that the learnt internal representation was 256 dimensional and c=2 would include only buy side and sell side spoofing instances while c=3 would include neutral instances along with buy side and sell side. We log the accuracy and F1 score in this order, i.e $(x,y)$ is the tuple of metrics listed below, where x represents the accuracy and y represents the F1 score. 

\vspace{0.5cm}
\setlength{\arrayrulewidth}{0.5mm}
\setlength{\tabcolsep}{10pt}
\renewcommand{\arraystretch}{1.5}
\begin{tabular}{ |p{1cm}|p{1cm}|p{1cm}|p{1cm}|  }
\hline
\multicolumn{2}{c|}{$E = 256$} & 
\multicolumn{2}{c}{$E = 1024$} \\
\hline
$c=2$ & $c=3$ & $c=2$ & $c=3$
 \\
\hline
(90.97, 80.40) & (86.02, 46.08) & (90.18, 56.90)
& (84.36, 45.85)\\
\hline
\end{tabular}
\vspace{0.1cm}
\label{tab:results}

\section{Conclusions and Outlook}
We have introduced a framework for detecting market manipulation and, more importantly, ranking potential spoofing activity based on expert input. We conducted a preliminary experiment to demonstrate the predictability of our target. We propose to use the learnt representations of the suspicious activity annotated by experts to rank future discoveries. Where experts are unavailable, we would employ a secondary algorithm to detect more complex scenarios of spoofing. For example, this algorithm would provide additional attention to multi deletion scenarios, top of the book spoof orders, and continuous patterns of spoofing. We will use this algorithm in place of an expert to test our hypothesis whether we can effectively identify and rank future complex scenarios of spoofing. To the best of our knowledge this is the first work exploring the idea of learning potential market order book abstractions to understand spoofing.

\bibliography{aaai24}

\begin{thebibliography}{6}
\providecommand{\natexlab}[1]{#1}

\bibitem[{CFTC(2019)}]{cftcfine}
CFTC. 2019.
\newblock Release Number 7946-19.

\bibitem[{ESMA(2018)}]{mar}
ESMA. 2018.
\newblock MAR.
\newblock Paper.
\newblock Regulation.

\bibitem[{Inc(2016)}]{tradingtech}
Inc, T. T.~I. 2016.
\newblock Applied Artificial Intelligence Technology For Processing Trade Data To Detect Patterns Indicative Of Potential Trade Spoofing.
\newblock USA.
\newblock ISBN US10552735B1.

\bibitem[{Leea, Eomb, and Parkd(2013)}]{Leeaun}
Leea, E.; Eomb, K.~S.; and Parkd, K.~S. 2013.
\newblock Microstructure-based manipulation: Strategic behavior and performance of spoofing traders.

\bibitem[{Qureshi(2019)}]{Qureshi}
Qureshi, F. 2019.
\newblock Stock Market Layering Fraud as a Stochastic Optimal Stopping Problem.
\newblock Presentation.

\bibitem[{Ramachandran, Zoph, and Le(2017)}]{ramach2017searching}
Ramachandran, P.; Zoph, B.; and Le, Q.~V. 2017.
\newblock Searching for Activation Functions.
\newblock arXiv:1710.05941.

\end{thebibliography}

\end{document}